\documentclass[12pt, aps, preprint]{revtex4}
\usepackage{float}
\usepackage{epsfig}
\usepackage{amssymb}
\usepackage{graphicx}%
\usepackage{amsmath}
\usepackage{color}
\usepackage{times}
\usepackage{verbatim}

\makeatletter
\def\btt#1{\texttt{\@backslashchar#1}}%
\DeclareRobustCommand\bblash{\btt{\@backslashchar}}%
\makeatother 
\begin{document}

%%%%%%%%%%%%%%%%%%%%%%%%%%%%%%%Start Kun%%%%%%%%%%%%%%%%%%%%%%%%%%%%
\begin{center}
{\Large\bf{ Impact of Stock Market Structure on \\Intertrade Time and Price Dynamics }}

\vspace{0.7cm} { Ainslie~Yuen$^{1}$ \& Plamen~Ch.~Ivanov$^{2}$}
\end{center}

\noindent{\small{\it{$^{1}$Signal Processing Laboratory, Department of Engineering, University of Cambridge, \\Cambridge~CB2~1PZ, UK\\
$^{2}$Center for Polymer Studies and Department of Physics,
  Boston University, Boston, MA 02215
}}}

%\centerline{\today}
\centerline{27 April, 2004}
%%%%%%%%%%%%%%%%%%%%%%%%%%%%%%%End Kun%%%%%%%%%%%%%%%%%%%%%%%%%%%%

{\bf Recent developments on the New York Stock Exchange (NYSE) \cite{WSJ_SEC,WSJ_SpecialistMan}, have raised the profile of the market operating mechanism, the ``market structure", employed by a stock market. The two major stock markets in the U.S., the NYSE and the National Association of Securities Dealers Automated Quotation System (NASDAQ) National Market have very different structures \cite{Hasbrouck93A,Smith98}, and there is continuing controversy over whether reported differences in stock price behaviour are due to differences in market structure or company characteristics \cite{NASDAQ_SEC_2001}. 
As the influence of market structure on stock prices may be obscured by exogenous factors such as demand and supply \cite{Gallant92,Lillo03},
we hypothesize that modulation of the flow of transactions due to market operations may carry a stronger imprint of the internal market mechanism. 
To this end, we analyse times between consecutive transactions for a diverse group of stocks registered on the NYSE and NASDAQ markets, and we relate the dynamical properties of the intertrade times with those of the corresponding price fluctuations. 
We report evidence of a robust scale-invariant temporal organisation in the transaction timing of stocks which is independent of individual company characteristics and industry sector, but which depends on market structure. 
Specifically, we find that 
stocks registered on the NASDAQ exhibit significantly stronger correlations in their transaction timing within a trading day, compared with NYSE stocks. Further, we find that companies that transfer from the NASDAQ to the NYSE show a reduction in the correlation strength of transaction timing within a trading day, after the move, suggesting influences of market structure. 
Surprisingly, we also observe that 
stronger power-law correlations in intertrade times are coupled with stronger power-law correlations in absolute price returns and higher price volatility, suggesting a strong link between
the dynamical properties of intertrade times and the corresponding price fluctuations over a broad range of time scales.
Comparing the NYSE and NASDAQ markets, 
we demonstrate that the higher correlations we find in intertrade times for NASDAQ stocks are indeed matched by higher correlations in absolute price returns and by higher volatility,
 suggesting that market structure may affect price behaviour through information contained in transaction timing.
These findings may have implications for the use of transaction timing in the prediction of prices and risk management on different stock markets.}

%%%%%%%%%%%%%%%%%%%%%%%%%%%%%%%%%%%%%%%%%%%%%%%%%%%%%%%%%%%%%%%%%%%%%%%%%%%%%%%%
The impact of market structure and associated rules of operation on market efficiency and stock 
price formation have attracted considerable public attention in recent months \cite{WSJ_SEC}.
This has also been of concern 
to those involved in stock market regulation, on behalf of investors \cite{SEC_NMS,WSJ_SEC}, since optimising market structure results in more effectively functioning markets \cite{Bennett03} and increases a market's competitiveness for market share in listed stocks \cite{Masulis02}.
Recent comparative studies of the NYSE and NASDAQ markets have primarily focused on stock prices to provide evidence that market organisational structure affects the price formation process \cite{Bessembinder97,Weaver02,Bennett03}.
Specifically, it has been shown that stocks registered on the 
NASDAQ may be characterised by a larger bid-ask spread \cite{Christie94A} and higher price volatility \cite{Bessembinder97,Weaver02,Bennett03}. 
However, this is often attributed to the market capitalisation, growth rate or the nature of the companies listed on the NASDAQ \cite{NASDAQ_SEC_2001}.
Furthermore, empirical studies have emphasized the dominant role and impact of trading volume on prices \cite{Gallant92,Lillo03}; since traded volume is determined by investors it is difficult to isolate the effects of market structure on price formation.

To better understand the mechanism by which market structure may affect stock prices, we study the 
information contained in the times between consecutive stock transactions. 
As market-specific operations may modulate the flow of transactions, we hypothesize that dynamical features of transaction timing reflect the underlying market mechanism.
Specifically, we ask if stocks of companies with diverse characteristics registered on a 
given market exhibit common features in their transaction timing, which may be associated with the particular 
market structure.
Further, we investigate how the dynamical properties of transaction timing relate to dynamical features of price fluctuations over a range of time scales, and whether market structure affects the temporal organisation of price fluctuations.

To probe how market structure influences the trading of stocks, we consider the two major U.S. stock 
markets, the NYSE and the NASDAQ.
All transactions on the NYSE of a given stock are centralised and 
are controlled by a {\it single} human operator called a ``specialist", whose primary role is to match together public buy and sell orders on the basis of price, in an auction-like setting \cite{Hasbrouck93A}. The NYSE specialist is under obligation to maintain both price continuity and a ``fair and orderly market" \cite{Hasbrouck93A}. 
The specialist is also under obligation to intervene, using his own firm's inventory of available stock, to provide liquidity in the event of an order imbalance, preventing sharp changes in the stock price \cite{Hasbrouck93A}.
The NYSE regulations allow for considerable flexibility within the specialist's operations \cite{WSJ_SpecialistMan}. 
In contrast, trading on the NASDAQ is decentralised, with trading in a given stock managed by a {\it number} of dealers called ``market makers".
Each market maker maintains his own inventory of stock in order to buy and sell to the public \cite{Smith98}. 
In comparison, the NYSE specialist rarely uses his own firm's inventory: such transactions involve less than 15\% of trading volume \cite{WSJ_Fidelity}. 
Although several regional exchanges may trade NYSE listed stocks, price formation has primarily been attributed to NYSE trading \cite{Hasbrouck95}. In contrast, the NASDAQ market relies on competition between multiple dealers for public orders to facilitate the price formation process \cite{Christie94A}. 
Moreoever, a substantial fraction of share volume on the NASDAQ is not handled by dealers, but is traded electronically via networks for small public orders and for institutional investors \cite{Smith98}. 
Such fragmentation of the NASDAQ stock market has been associated with higher price volatility \cite{Bennett03}.

Here we ask to what extent such structural and operational differences between the NYSE and NASDAQ markets affect the flow of transactions. 
It is difficult to answer whether differences in intertrade times are due to individual company characteristics because the majority of empirical studies have considered a single stock over a few months \cite{Engle98,Raberto02}. Studies which considered the intertrade times of a larger group of stocks did not find common features in the intertrade times and did not compare between markets \cite{Lo92,Dufour00}.
The only comparative study considered a single NYSE and a single Paris stock, finding some differences in their intertrade times, but those may well be due to a different culture of trading \cite{Jasiak99}. 
To probe for evidence of the impact of market structure on the trading of stocks, we employ concepts and methods from statistical physics to investigate the correlation properties of transaction timing for diverse companies, over time scales ranging from seconds up to a year.

We examine one hundred stocks listed on the NYSE, from eleven industry sectors: Technology-Hardware(5), Semiconductors(2), Pharmaceutical \& Medical Equipment(10), Financial(8), Automotive(9), Defense/Aerospace(9), Mining, Metals \& Steel Works(8), Chemicals \& Plastics(7), Retail \& Food(17), Petroleum, Gas \& Heavy Machinery(10), Telephone Service Providers(7), Electric \& Power Services(8). We study the time intervals between successive stock trades, over a period of four years - 4 Jan. 1993 to 31 Dec. 1996 - as recorded in the Trades and Quotes (TAQ) database from the NYSE (Table~\ref{Table 1}). 
We also analyse one hundred NASDAQ stocks from fourteen industry sectors: Technology-Hardware(28), Technology-Software(16), Semiconductors(7), Pharmaceutical, Biotechnology \& Medical Equipment(12), Financial(5), Automotive(1), Steel Works(1), Chemicals(1), Retail \& Food(16), Petroleum, Gas \& Heavy Machinery (2), Telephone \& Cable Television Service Providers(5), Services(2),  Transportation(3), Electrical Apparatus(1). We study the time intervals between successive stock trades as recorded in the TAQ database, for twenty-nine companies over the period 4 Jan. 1993 - 31 Dec. 1996, and seventy one companies  over the period 3 Jan. 1994 - 30 Nov. 1995 (marked with (*) in Table~\ref{Table 2}). 
For both markets, we select companies with average market capitalisations ranging over three decades, and varying levels of
trading activity with average values of intertrade time between 11 and 640 seconds for NYSE stocks, and between 5 and 680 seconds for NASDAQ stocks.
In parallel with the intertrade times, we analyse the prices for both sets of stocks over the same periods.

Like many financial time series the intertrade times (ITT) are inhomogeneous and nonstationary, with statistical properties changing with time, e.g. ITT data exhibit trends superposed on a pattern of daily activity. While ITT fluctuate in an irregular and complex manner on a trade-by-trade basis, empirical observations reveal that periods of inactive trading are often followed by periods of more active trading (Fig.~\ref{fig.1}). Such patterns can be seen at scales of observation ranging from minutes to months, suggesting that there may be a self-similar, fractal structure in the temporal organisation of intertrade times, independent of the average level of trading activity of a given stock.

To probe for scale-invariant features in the fluctuations of intertrade times, we apply the detrended fluctuation analysis (DFA) method, which has been shown to detect and accurately quantify long-range power-law correlations embedded in noisy non-stationary time series with polynomial trends \cite{Peng94}. We choose this method because traditional techniques such as power spectral, autocorrelation and Hurst analyses are not suited to nonstationary data \cite{Taqqu95}.
The DFA method (DFA-$l$) quantifies the root-mean-square fluctuations $F(n)$ of a signal at different time scales $n$, after accounting for nonstationarity in the data by subtracting underlying polynomial trends of order ($l-1$). A power-law functional form $F(n) \sim n^{\alpha}$ indicates self-similarity and fractal scaling in the ITT time series. The scaling exponent $\alpha$ quantifies the strength of correlations in the ITT fluctuations: if $\alpha = 0.5$ there are no correlations, and the signal is uncorrelated random noise; if $\alpha < 0.5$ the signal is anti-correlated, meaning that large values are more likely to be followed by small values; if $\alpha > 0.5$ there are positive correlations and the signal exhibits persistent behaviour, where large values are more likely to be followed by large values and small values by small values. The higher the value of $\alpha$, the stronger the correlations. The DFA method avoids the spurious detection of apparent long-range correlations that are an artifact of polynomial trends and other types of nonstationarities \cite{Taqqu95,Hu01,Kantelhardt01}.

We find that the ITT series for all stocks on both markets exhibit long-range power-law correlations over a broad range of time scales, from several trades to hundreds of thousands of trades, characterised by a scaling exponent $\alpha > 0.5$ (Fig.~\ref{fig.2}). For all stocks on both markets we observe a crossover in the scaling curve $F(n)$ from a scaling regime with a lower exponent $\alpha_{1}$ over time scales less than a trading day, to a scaling regime with an exponent $\alpha_{2} > \alpha_{1}$ (stronger positive correlations) over time scales from days to almost a year.

Further, we find that this crossover is systematically more pronounced for NYSE stocks compared with NASDAQ stocks. Characterising ITT fluctuations over time scales less than a day, we find that NASDAQ stocks exhibit systematically stronger correlations than NYSE stocks, with statistically significantly higher average value of the exponent $\alpha_{1}^{ITT_{NASDAQ}}=0.75 \pm 0.04$ (group mean $\pm$ std. dev.)
 as compared with $\alpha_{1}^{ITT_{NYSE}}=0.62 \pm 0.03$ (Fig.~\ref{fig.2}). In contrast, over time horizons above a trading day, we find that the correlation properties of ITT on both markets are statistically similar with average scaling exponent $\alpha_{2}^{ITT_{NASDAQ}}=0.85 \pm 0.08$ comparable with   $\alpha_{2}^{ITT_{NYSE}}=0.87 \pm 0.09$ (Fig.~\ref{fig.2}).

We next investigate how the correlation properties of ITT depend on the average level of trading activity, and if this dependence differs with market structure.
Since both sets of a hundred stocks that we study on the NYSE and NASDAQ markets encompass a wide range of average trading activity, we split both sets into six subsets with matching average ITT ($\overline{ITT}$) and approximately equal numbers of stocks (Fig.~\ref{fig.2}b,c). 
Within each market we find that over time scales less than a day, the correlation exponent characterising the trading dynamics is larger for stocks with higher trading activity (lower $\overline{ITT}$) (Fig.~\ref{fig.3}a).
Surprisingly, this dependence persists over much longer time scales, ranging from days to months (Fig.~\ref{fig.3}b). For NYSE stocks we find a logarithmic dependence of $\alpha_{1}^{ITT}$ and $\alpha_{2}^{ITT}$ on $\overline{ITT}$.
We then compare the scaling behaviour of ITT for each subset of NASDAQ stocks with the subset of NYSE stocks with matching $\overline{ITT}$.
We find that the average correlation exponent $\alpha_{1}^{ITT}$ for the NASDAQ stocks is consistently higher compared with the NYSE stocks for each subset, and that the difference $\alpha_{1}^{ITT_{NASDAQ}}-\alpha_{1}^{ITT_{NYSE}}$ is practically independent of $\overline{ITT}$ (Fig.~\ref{fig.3}a). In contrast, there is no systematic difference in the values of the average $\alpha_{2}^{ITT}$ for NASDAQ and NYSE stocks for subsets with matching $\overline{ITT}$ (Fig.~\ref{fig.3}b). These observations suggest that the difference in the correlation properties of intertrade times of NYSE and NASDAQ stocks is independent of the average level of trading activity.

Since for both NYSE and NASDAQ stocks we have chosen a range of market capitalisations, industry sectors and average levels of trading activity, our findings of a crossover in the scaling behaviour of ITT and stronger correlations over intraday time scales for NASDAQ stocks, support our hypothesis that market structure affects the dynamics of transaction timing.
However, more established companies listed on the NYSE may be subject to different trading patterns when compared with the younger and more rapidly growing companies on the NASDAQ. To verify that the stronger correlations in ITT over time scales less than a day for NASDAQ stocks are indeed due to market structure, we ask if the scaling properties of ITT systematically change for companies that transfer from the NASDAQ to the NYSE. In particular, we investigate the trading dynamics of ten companies that moved from the NASDAQ to the NYSE around the end of 1994 and the beginning of 1995 (Table~\ref{Table 3}). For each company, we analyse the ITT time series while the company was on the NASDAQ and then repeat the analysis when the company was on the NYSE.

For all ten companies we find a significant change in the scaling properties of intertrade times: a marked decrease in the strength of the power-law correlations within a trading day (lower $\alpha_{1}^{ITT}$) associated with the transfer from the NASDAQ to the NYSE (Fig.~\ref{fig.4}). There is however, no corresponding systematic change in the correlations over time scales above a trading day, consistent with our findings of statistically similar values of scaling exponent $\alpha_{2}^{ITT}$ for the two groups of one hundred stocks registered on the NYSE and NASDAQ (Fig.~\ref{fig.2}).
Thus, our results indicate that market structure impacts not only trading dynamics on a trade-by-trade basis \cite{Dufour00}, but also the fractal temporal organisation of trades
over time scales up to a day. The presence of stronger intraday correlations in transaction timing for NASDAQ stocks may be attributed to the multiplicity of dealers (ranging from 2 to 50 per stock during 1994 \cite{Christie94A}) and electronic methods of trading (Electronic Communication Networks and the Small Order Execution System \cite{Smith98}), allowing the NASDAQ to efficiently absorb fluctuations in trading activity in almost real time \cite{Masulis02}.
In contrast, for each stock on the NYSE, while there is the electronic SuperDOT routing system, each order has to be exposed to and compared with outstanding orders, as the single NYSE specialist finds the best bid to match an offer with \cite{Hasbrouck93A}. This may lead to interruptions in the execution of a rapid succession of trades on the NYSE, resulting in weaker correlations in intertrade times within a trading day.

On the other hand, our finding of stronger power-law correlations for both markets over time horizons from a trading day to several months ($\alpha_{2}^{ITT} > \alpha_{1}^{ITT}$) suggests that investors' behaviour is more coherent over longer time scales, as information driving trading activity takes time to disseminate.
Moreover, this can account for the similar values of $\alpha_{2}^{ITT}$ for subsets of NYSE and NASDAQ stocks with matched $\overline{ITT}$, since news and information driving trading activity are exogenous to market structure.

Finally, we investigate if the market-mediated differences in long-range power-law correlations in ITT translate into differences in the scaling behaviour of price fluctuations of stocks registered on the NASDAQ and NYSE markets.
To this end, in parallel with ITT we analyse the absolute price returns for each company in our database for both markets. 
For all stocks we observe a crossover at a trading day in the scaling function $F(n)$ of price fluctuations \cite{Liu99}, from weaker to stronger correlations, corresponding to the crossover we observe for intertrade times.
In addition we find that over time scales less than a day, stocks with stronger correlations in ITT exhibit stronger correlations in absolute price returns.
In particular, we find that the stronger correlations in ITT associated with the NASDAQ market structure ($\alpha_{1}^{ITT_{NASDAQ}} > \alpha_{1}^{ITT_{NYSE}}$), are accompanied by stronger correlations in price fluctuations ($\alpha_{1}^{|RET|_{NASDAQ}} > \alpha_{1}^{|RET|_{NYSE}}$) over time scales within a trading day (Fig.~\ref{fig.5}a).
We also find evidence of a positive relationship between correlations in ITT and correlations in price fluctuations over time scales larger than a trading day for NASDAQ stocks. In contrast, there is no corresponding positive relationship for NYSE stocks, suggesting a weaker coupling between trading dynamics and price formation under the NYSE market structure, over time horizons above a trading day. 
While previous work has suggested that bursts of trading activity have an instantaneous impact on stock prices \cite{Dufour00,Jones94}, our results show that the interaction between trading times and price formation may be more complex, where fractal temporal patterns in ITT are linked with scaling features of price fluctuations over a broad range of time scales.

We then test whether long-range correlations in ITT may be linked with stock price volatility. Previous studies have reported higher price volatility for NASDAQ stocks compared with NYSE stocks \cite{Bessembinder97,Weaver02,Bennett03}.
We find a positive relationship, with stronger correlations in ITT over time scales less than a day related to higher daily volatility $\sigma^{RET}$, and that the NASDAQ stocks have higher $\alpha_{1}^{ITT}$ and correspondingly higher $\sigma^{RET}$ compared with NYSE stocks (Fig.~\ref{fig.5}b).
This relationship may appear to follow from our observation that $\alpha_{1}^{ITT}$ depends on $\overline{ITT}$ (Fig.~\ref{fig.3}a), and previous studies which connect price volatility with periods of high transaction rates \cite{Engle98}. However, for the stocks in our database (Tables~\ref{Table 1} \& \ref{Table 2}), we find no clear dependence between $\sigma^{RET}$ and average level of trading activity as measured by $\overline{ITT}$ (Fig.~\ref{fig.5}c). Thus the relationship between $\alpha_{1}^{ITT}$ and $\sigma^{RET}$ suggests that information contained in the microscopic temporal structure of ITT is carried over a range of scales to impact daily price volatility.

In summary, our results indicate that market structure influences the correlation properties of transaction timing, with stocks registered on the NASDAQ showing systematically stronger long-range, power-law correlations within a trading day compared with stocks listed on the NYSE. 
Moreover, stocks characterised by stronger correlations in their intertrade times exhibit stronger correlations in their price fluctuations and higher daily price volatility. Further, stocks registered on the NASDAQ are characterised not only by higher volatility compared with NYSE stocks, but also by stronger correlations in price fluctuations over time scales less than a day, suggesting an influence of market structure on the process of price formation over a range of time scales.
Understanding the scale-invariant properties of intertrade times is crucial to the development of more realistic models of the price formation process \cite{Lo92,Ghysels95A,Mandelbrot97,Engle98,Masoliver03} and its dependence on market structure.
Furthermore, these results are of interest in the context of the continuing process of optimising market structure to maintain the efficiency and competitiveness of U. S. stock markets.

%%%%%%%%%%%%%%%%%%%%%%%%%%%%%%%%%%%%%%%%%%%%%%%%%%%%%%%%%%%%%%%%%%%%%%
\bibliography{econ_bibfile}

%%%%%%%%%%%%%%%%%%%%%%%%%%%%%%%%%%%%%%%%%%%%%%%%%%%%%%%%%%%%%%%%%%%%%%

%%%%%%%%%%%%%%%%%%%%%%%%%%%%%%%%%%%%%%%%%%%%%%%%%%%%%%%%%%%%%%%%%%%%%%%%%%%%%%%%
%   Tables
%%%%%%%%%%%%%%%%%%%%%%%%%%%%%%%%%%%%%%%%%%%%%%%%%%%%%%%%%%%%%%%%%%%%%%%%%%%%%%

\begingroup
\squeezetable
\begin{table*}
\caption{ }
\label{Table 1}
\begin{centering}
\begin{tabular}{crcrr|crcrr} \hline
Company & Ticker & Industry & Number & $\overline{ITT}$  &  Company & Ticker & Industry & Number & $\overline{ITT}$ \\

Name & Symbol &  & of Trades & (sec) &  Name & Symbol &  & of Trades & (sec) \\
\hline

Meredith & MDP & Food \& Retail  & 35267 & 636 & Medtronics &MDT & Medical Apparatus & 308049& 75\\ [-2.5pt]
Transco & E   &   Natural Gas &   47045      &   405    & Southern & SO   &   Electric Services &   329464&   71   \\ [-2.5pt] 
Avery Dennison & AVY   &   Paper Products &   62927      &   365   & Schlumberger & SLB   &   Oil \& Gas      &   330830&   70   \\ [-2.5pt] 
Johnson Controls & JCI   &   Automatic Controls     &   68490      &   334    & Amoco & AN   &   Petroleum     &   339996&   69   \\ [-2.5pt] 
Northrop Grumman & NOC   &   Aerospace/Defense     &   69739      &   330    & PG \& E & PCG   &   Electric Services     &   355190&   66   \\ [-2.5pt] 
Allergan & AGN   &   Pharmaceutical     &   71419      &   322    & Sprint PCS & FON   &   Telephone Comms.     &   362851&   64   \\ [-2.5pt] 
Jefferson Pilot & JP   &   Financial     &   79013      &   292    & Homestake Mining & HM   &   Mining     &   370132&   63   \\ [-2.5pt] 
Nalco Chemical & NLC   &   Chemicals     &   81731      &   283    & Union Carbide & UK   &   Chemicals     &   387273&   60   \\ [-2.5pt] 
Lockheed Martin & LK   &   Aerospace/Defense     &   44897      &   282    & Nynex & NYN   &   Telephone Comms.     &   386703&   60   \\ [-2.5pt] 
Northern States Pow. & NSP   &   Electric Services     &   85724      &   269    & Morgan J.P. \& Co. & JPM   &   Financial     &   401213&   58   \\ [-2.5pt] 
Dana & DCN   &   Automotive     &   89700      &   257    & Dow Chemical & DOW   &   Chemicals    &   411258&   57   \\ [-2.5pt] 
Inland Steel Ind. & IAD   &   Steelworks     &   91137      &   253    & Mobil & MOB   &   Petroleum Refining     &   430401&   54   \\ [-2.5pt] 
Ashland Inc. & ASH   &   Petroleum Refining     &   94396      &   245    &  Schering Plough & SGP   &  Pharmaceutical     &   431388&   54   \\ [-2.5pt] 
General Dynamics & GD   &   Aerospace/Defense     &   97594      &   237     &  Chase Manhattan & CMB   &   Financial     &   448801&   52   \\ [-2.5pt] 
Eaton & ETN   &   Automotive     &   98796      &   234    & BellSouth & BLS   &   Telephone Comms.     &   450144&   52   \\ [-2.5pt] 
Ethyl & EY   &   Chemicals     &   100663      &   229    & 3M & MMM   &   Paper Products    &   449462&   52   \\ [-2.5pt] 
TRW Inc. & TRW   &   Automotive     &   111506      &   208    & Texaco & TX   &   Petroleum     &   457081&   51   \\ [-2.5pt] 
 Alcan Aluminium &   AL   &   Metals     &   112193      &   207    &  Arch. Dan. Midl. &  ADM   &   Food     &   468148&   50   \\ [-2.5pt] 
 Unilever &   UN   &   Food \& Retail   &   113736      &   203    &   Bell Atlantic & BEL   &   Telephone Comms.     &   499768&   47   \\ [-2.5pt] 
Union Electric &  UEP   &   Electric Services     &   119737      &   193    &  Pacific Telesis & PAC   &   Telephone Comms.    &   508091&   46   \\ [-2.5pt] 
Hercules &  HPC   &   Chemicals     &   123618      &   187    &   Lilly Eli \& Co. & LLY   &   Pharmaceutical    &   514899&   45   \\ [-2.5pt] 
 Air Prod. \& Chem. & APD   &   Chemicals    &   123416      &   187    &  Sara Lee &  SLE   &   Food \& Retail     &   527814&   44   \\ [-2.5pt] 
  Textron & TXT   &   Aerospace/Defense    &   123879      &   187    &    Dupont & DD   &   Chemicals     &   543724&   43   \\ [-2.5pt] 
Carolina Power\&Light &  CPL   &   Electric Services     &   131352      &   177    &  American Express & AXP   &   Financial   &   581840&   40   \\ [-2.5pt] 
 Nortel Networks &   NT   &   Telephone Apparatus    &   132384      &   176    &   Fed. Nat. Mort. & FNM   &   Financial    &   627313&   37   \\ [-2.5pt] 
Baltimore Gas \& Elec. &  BGE   &   Electric Services     &   142973      &   163    &  Adv. Micro Dev. &  AMD   &   Semiconductors     &   644865&   36   \\ [-2.5pt] 
  Hershey Foods & HSY   &   Food \& Retail     &   144982      &   160    &   Citicorp & CCI   &   Financial     &   677484&   34   \\ [-2.5pt] 
 Honeywell Int. &  HON   &   Aerospace/Defense    &   156376      &   149    &   Abbott Labs. & ABT   &   Pharmaceutical     &   691877&   34   \\ [-2.5pt] 
Navistar Int. &  NAV   &   Automotive     &   168951      &   138    &   Pfizer & PFE   &   Pharmaceutical    &   689705&   34   \\ [-2.5pt] 
Campbell Soup &  CPB   &  Food \& Retail     &   175869  &  132  & Texas Instruments &  TXN   & Semiconductors  & 708329& 33 \\ [-2.5pt] 
Raytheon &   RTN   &   Aerospace/Defense     &   176148      &   132    &    Boeing Aerospace & BA   &   Aerospace/Defense    &   728779&   32   \\ [-2.5pt] 
United Tech. &   UTX   &   Aerospace/Defense     &   190049      &   122    &   Exxon & XON   &   Petroleum Refining     &   750298&   31   \\ [-2.5pt] 
Nucor &  NUE   &   Steelworks    &   194532      &   119    &   Johnson \& Johnson & JNJ   &   Pharmaceutical    &   1001549&   23   \\ [-2.5pt] 
Barnett Banks &   BBI   &   Financial     &   202774      &   115    &  Hewlett-Packard &  HWP   &   Hardware     &   1094829&   21   \\ [-2.5pt] 
Phelps Dodge &   PD   &   Metal Refining     &   203834      &   114    &    Home Depot & HD   &   Retail    &   1103037&   21   \\ [-2.5pt] 
McDonnell Douglas &    MD   &   Aerospace/Defense   &   203845      &   114    &  Brist. Myers Squibb &  BMY   &   Pharmaceutical    &   1121714&   21   \\ [-2.5pt] 
Fluor &  FLR   &   Construction     &   205913      &   113    &   General Motors & GM   &   Automotive     &   1130452&   21   \\ [-2.5pt] 
General Mills &  GIS   &   Food \& Retail    &   227318      &   103    &   Compaq Computer & CPQ   &   Hardware   &   1184985&   20   \\ [-2.5pt] 
Newmont Mining &  NEM   &   Mining    &   232391      &   100    &  Chrysler &    C   &   Automotive    &   1231979&   19   \\ [-2.5pt] 
Anheuser Busch & BUD   &   Food \& Retail    &   251972      &   93    &    Coca Cola & KO   &   Food \& Retail     &   1244660&   19   \\ [-2.5pt] 
 USX-US Steel Grp. & X   &   SteelWorks   &   252435      &   92    &     Ford & F   &   Automotive   &   1260730&   19   \\ [-2.5pt] 
Alza &  AZA   &   Pharmaceutical     &   257116      &   91    &   GTE & GTE   &   Telephone Comms.    &   1268523&   18   \\ [-2.5pt] 
Alcoa &   AA   &   Metal Refining  &   260980      &   89    &   Pepsico & PEP   &   Food \& Retail  &   1321427&   18   \\ [-2.5pt] 
Bank Boston &  BKB   &   Financial  &   262506      &   89    &   General Electric & GE   &   Food \& Retail  &   1374682&   17   \\ [-2.5pt] 
Colgate Palmolive &   CL   &   Food \& Retail  &   262896      &   88    &    Philip Morris & MO   &   Food \& Retail   &   1527659&   15   \\ [-2.5pt] 
 Goodyear Tire \& Rub. &   GT   &   Automotive  &   272025      &   85    & IBM &  IBM   &   Hardware  &   1677319&   14   \\ [-2.5pt] 
Niagara Mohawk Pow. &  NMK   &   Electric Services   &   276284      &   84    &   AT\&T &  T   &   Telephone Comms.  &   1689767&   14   \\ [-2.5pt] 
Atlantic Richfield &  ARC   &   Petroleum    &   286580      &   81    &   Wal Mart & WMT   &   Retail   &   1794160&   13   \\ [-2.5pt] 
FPL Group &  FPL   &   Electric Services  &   303364      &   77    &   Merck \& Co. & MRK   &   Pharmaceutical   &   2055443&   11   \\ [-2.5pt] 
Royal Dutch Petrol. &   RD   &   Petroleum  &   304505      &   76    & Motorola &  MOT   &   Hardware  &   2204059&   11   \\ 

\hline
 \end{tabular}
\end{centering}
\end{table*}

\endgroup

\clearpage
\noindent Table \ref{Table 1}: Characteristics of one hundred NYSE stocks studied over the period 4 Jan. 1993 - 31 Dec. 1996. Companies range in average market capitalisation from $\$0.8 \times 10^{9}$ to $\$102 \times 10^{9}$ over the period, and are ranked in order of decreasing average value of ITT ($\overline{ITT}$). We include all trades occurring during NYSE trading hours (9.30am-4pm EST), excluding public holidays and weekends.
\clearpage

\begingroup
\squeezetable
\begin{table*}
\caption{ }
\label{Table 2}
\begin{centering}
\begin{tabular}{cccrr|cccrr} \hline
Company & Ticker & Industry & Number & $\overline{ITT}$ &  Company & Ticker & Industry & Number & $\overline{ITT}$ \\ 

Name & Symbol &  & of Trades & (sec) &  Name & Symbol & & of Trades & (sec) \\
\hline

Oshkosh B Gosh & GOSHA   &   Retail \& Food     &   31986&   683   & US Robotics & USRX*   &   Hardware     &   143912&   78   \\ [-2.5pt] 
Sanmina-SCI & SANM*   &   Hardware     &   22648&   438   & Symantec & SYMC*   &   Software     &   143405&   77   \\ [-2.5pt] 
MedImmune &  MEDI*   &   Biotech.     &   24618      &   414   & Autodesk & ACAD   &   Software     &   261716&   74   \\ [-2.5pt] 
ICOS & ICOS*   &   Pharmaceutical     &   32460      &   339   & Oxf. Health Plans & OXHP*   &   Financial     &   150501&   74   \\ [-2.5pt] 
Gilead Sciences &  GILD*   &   Biotech.     &   32187      &   332   & Komag & KMAG*   &   Hardware     &   175953&   63   \\ [-2.5pt] 
Molex & MOLX*   &   Hardware     &   34104      &   321   & Biomet & BMET   &   Med. Apparatus     &   379342&   62   \\ [-2.5pt] 
Coors Adolph & ACCOB   &   Food \& Retail     &   78393&   295   & Novellus Systems & NVLS*   &   Hardware     &   182185&   61   \\ [-2.5pt] 
Whole Foods Mar. & WFMI*   &   Food \& Retail     &   38018      &   290   & Mobile Tel. Tech. & MTEL*   &   Telephone Comms.     &   184469&   61   \\ [-2.5pt] 
Ross Stores &  ROST*   &   Food \& Retail     &   42772      &   256   & KLA-Tencor & KLAC*   &   Hardware     &   187614&   60   \\ [-2.5pt] 
XOMA & XOMA*   &   Pharmaceutical     &   43073      &   256   & St. Jude Medical & STJM   &   Med. Apparatus     &   388393&   59   \\ [-2.5pt] 
Paccar & PCAR   &   Automotive     &   94496&   245   & AST Research & ASTA*   &   Hardware     &   191683&   58   \\ [-2.5pt] 
General Nutr. Cos. & GNCI*   &   Food \& Retail     &   48222      &   226   & Parametric Tech. & PMTC*   &   Software     &   197637&   57   \\ [-2.5pt] 
Ryans Fam. Steak. & RYAN   &   Food \& Retail     &   108243&   215   & Starbucks & SBUX*   &   Food \& Retail     &   201225&   56   \\ [-2.5pt] 
Caliber System &  ROAD   &   Transportation     &   106570&   209   & Read-Rite & RDRT*   &   Hardware     &   205021&   54   \\ [-2.5pt] 
Giddings \& Lewis & GIDL   &   Heavy Machinery     &   57081      &   204   & Borland Software & BORL*   &   Software     &   207697&   54   \\ [-2.5pt] 
Huntington Banc. & HBAN*   &   Financial     &   55885      &   199   & Gateway 2000 & GATE*   &   Hardware     &   217267&   52   \\ [-2.5pt] 
Worthington Ind. & WTHG   &   Steelworks     &   119751&   194   & LM Ericsson Tel. & ERICY*   &   Hardware     &   228287&   49   \\ [-2.5pt] 
Phycor &  PHYC*   &   Office Services     &   59431      &   183   & StrataCom &  STRM*   &   Hardware     &   235537&   48   \\ [-2.5pt] 
Intergraph & INGR   &   Hardware/Software     &   131780&   176   & Xilinx & XLNX*   &   Semiconductors     &   239423&   47   \\ [-2.5pt] 
Shared Medical Sys. & SMED   &    Hardware/Software     &   132579&   175   & Biogen & BGEN*   &   Biotech.     &   241886&   46   \\ [-2.5pt] 
Glenayre Tech. & GEMS*   &   Hardware     &   63152      &   174   & Adaptec & ADPT*   &   Hardware     &   253082&   44   \\ [-2.5pt] 
PETsMART & PETM*   &   Food \& Retail     &   67047      &   165   & Acclaim Ent. & AKLM*   &   Software     &   282481&   40   \\ [-2.5pt] 
Tyson Foods & TYSNA*   &   Food \& Retail     &   70711      &   158   & Chiron & CHIR*   &   Pharmaceutical     &   292353&   38   \\ [-2.5pt] 
MFS Comms. & MFST*   &   Telephone Comms.     &   70776      &   157   & Tellabs & TLAB*   &   Hardware     &   299490&   38   \\ [-2.5pt] 
Brunos & BRNO   &    Food \& Retail     &   99211&   155   & Adobe Systems & ADBE*   &   Software     &   307959&   36   \\ [-2.5pt] 
Sigma-Aldrich & SIAL*   &   Chemicals     &   72843      &   153   & America Online & AMER*   &   Services     &   314541&   36   \\ [-2.5pt] 
Atlantic S.E. Air. & ASAI*   &   Transportation     &   75031      &   148   & Electronic Arts & ERTS*   &   Software     &   329541&   34   \\ [-2.5pt] 
Cephalon & CEPH*   &   Pharmaceutical     &   71733      &   148   & Qualcomm & QCOM*   &   Hardware     &   335494&   34   \\ [-2.5pt] 
Safeco & SAFC   &   Financial     &   157461&   148   & Informix & IFMX*   &   Software     &   350185&   32   \\ [-2.5pt] 
Comcast & CMCSA   &   Cable TV     &   161408&   144   & Altera & ALTR*   &   Semiconductors     &   349925&   32   \\ [-2.5pt] 
Stew. \& Stev. Svcs &  SSSS*   &   Heavy Machinery     &   79177      &   141   & Tele Comms. & TCOMA   &   Cable TV     &   765301&   31   \\ [-2.5pt] 
American Greetings & AGREA   &   Food \& Retail     &   169265&   138   & Amer. Pow. Conv. & APCC*   &   Electrical Apparatus     &   395510&   28   \\ [-2.5pt] 
Northwest Airlines &  NWAC*   &   Transportation     &   77658      &   127   & Lotus Devel. & LOTS   &   Software     &   582256&   25   \\ [-2.5pt] 
ADC TeleComms. & ADCT*   &   Hardware     &   90573      &   123   & Integr. Dev. Tech. & IDTI*   &   Semiconductors     &   471169&   24   \\ [-2.5pt] 
Charming Shoppes & CHRS   &   Food \& Retail     &   196473&   119   & Cirrus Logic & CRUS*   &   Semiconductors     &   500710&   22   \\ [-2.5pt] 
HBO \& Co. & HBOC*   &   Hardware/Software      &   95662      &   116   & US HealthCare & USHC*   &   Financial     &   505215&   22   \\ [-2.5pt] 
Microchip Tech. & MCHP*   &   Semiconductors     &   102625      &   109   & MCI Comms. & MCIC   &   Telephone Comms.     &   1096316&   21   \\ [-2.5pt] 
Andrew & ANDW   &   Hardware     &   215063&   109   & DELL & DELL*   &   Hardware     &   557195&   20   \\ [-2.5pt] 
Legent & LGNT*   &   Software     &   90705      &   108   & DSC Comms. & DIGI   &   Hardware     &   1209063&   19   \\ [-2.5pt] 
Stryker & STRY*   &   Medical Apparatus     &   107678      &   104   & Applied Materials & AMAT*   &  Hardware     &   584276&   19   \\ [-2.5pt] 
PeopleSoft & PSFT*   &   Software     &   108433      &   102   & Sybase & SYBS*   &   Software     &   631753&   18   \\ [-2.5pt] 
Outback Steak. & OSSI*   &   Food \& Retail      &   112607      &   99   & Amgen & AMGN   &   Biotech.     &   1392229&   17   \\ [-2.5pt] 
Boatmens Banc. & BOAT   &   Financial     &   236139&   99   & 3Com & COMS*   &   Hardware     &   699889&   16   \\ [-2.5pt] 
Intelligent Elec. & INEL*   &   Hardware     &   113666      &   98   & Apple Computer & AAPL   &   Hardware     &   1646925&   14   \\ [-2.5pt] 
Genzyme General & GENZ*   &    Biotech.     &   116223      &   96   & Novell & NOVL   &   Software    &   1803407&   13   \\ [-2.5pt] 
Bed Bath \& Beyond & BBBY*   &   Food \& Retail     &   120723      &   92   & Oracle & ORCL   &   Software    &   1817365&   13   \\ [-2.5pt] 
Intuit & INTU*   &   Software     &   122051      &   91   & Sun Microsystems & SUNW   &   Hardware/Software     &   2029156&   12   \\ [-2.5pt] 
Boston Chicken & BOST*   &   Food \& Retail     &   128376      &   87   & Cisco Systems & CSCO*   &   Hardware     &   1093386&   10   \\ [-2.5pt] 
Staples &  SPLS*   &   Food \& Retail     &   132041      &   85   & Microsoft & MSFT*   &   Software    &   1505531&   7   \\ [-2.5pt] 
Linear Tech. & LLTC*   &   Semiconductors     &   139953      &   80   & Intel & INTC   &   Semiconductors    &   4807756&   5   \\ 

\hline
\end{tabular}
\end{centering}
\end{table*}

\clearpage
\noindent Table \ref{Table 2}: Characteristics of one hundred NASDAQ stocks studied; data covers twenty nine companies over the period 4 Jan. 1993 - 31 Dec. 1996, and seventy one companies (marked with~*~) over the period 3 Jan. 1994 - 30 Nov. 1995. Companies range in average market capitalisation from $\$0.2 \times 10^{9}$ to $\$40 \times 10^{9}$, and are ranked in order of decreasing average value of ITT ($\overline{ITT}$). We include all trades occurring during regular NASDAQ trading hours (9.30am-4pm EST), excluding public holidays and weekends.
\clearpage

\begingroup
\squeezetable
\begin{table*}
\caption{ Characteristics of ten stocks that moved from the NASDAQ to the NYSE during the period 3 Jan. 1994 - 30 Nov. 1995. Companies are ranked in order of decreasing average value of ITT when on the NYSE. We include all trades occurring during NYSE trading hours (9.30am-4pm EST) excluding public holidays and weekends. }
\label{Table 3}
\begin{centering}
\begin{tabular}{lc|ccrr|ccrr} \hline
Company & Industry &  \multicolumn{4}{c}{NASDAQ} & \multicolumn{4}{c}{NYSE} \\ \hline
  &   & Ticker & Number  & Number & $\overline{ITT}$ & Ticker  & Number & Number & $\overline{ITT}$\\
&   &  Symbol & of Days & of Trades & (sec) & Symbol & of Days & of Trades & (sec)\\

Input Output &  Measuring Devices &IPOP & 219 & 25211 & 198  & IO & 265 & 10944 & 540 \\ [-2.5pt]
Consolidated Papers &  Paper Mills &CPER & 154 & 8902 & 389 & CDP & 330 & 15180 & 488 \\ [-2.5pt]
Cardinal Health &  Wholesale Drugs &CDIC & 171 & 11510 & 333 & CAH & 313 & 14819 & 475 \\ [-2.5pt]
AK Steel Holding Corp. &  Steelworks &AKST & 256 & 14575 & 383 & AKS & 167 & 10397 & 364 \\ [-2.5pt]
Sports \& Recreation &  Retail &SPRC & 177 & 17721 & 222 & WON & 307 & 19907 & 345 \\ [-2.5pt]
State Street Boston &  Financial &STBK & 282 & 43829 & 148 & STT & 202 & 16916 & 273 \\ [-2.5pt]
Dollar General & Retail &DOLR & 273 & 34873 & 180 & DG & 211 & 19817 & 241  \\ [-2.5pt]
Mid-Atlantic Medical Services & Financial &MAMS & 187 & 90598 & 48 &  MME & 297 & 50245 & 136\\ [-2.5pt]
Seagate &  Hardware  &SGAT & 238 & 119544 & 46 & SEG & 246 & 85100 & 67 \\ [-2.5pt]
Newbridge Networks &  Hardware &NNCXF & 176 & 208771 & 20 & NN & 308 & 148637 & 28 \\ \hline
\end{tabular}
\end{centering}
\end{table*}

%%%%%%%%%%%%%%%%%%%%%%%%%%%%%%%%%%%%%%%%%%%%%%%%%%%%%%%%%%%%%%%%%%%%%%%%%%%%%%%%

%%%%%%%%%%%%%%%%%%%%%%%%%%%%%%%%%%%%%%%%%%%%%%%%%%%%%%%%%%%%%%%%%%%%%%%%%%%%%%%%
%   Figure Captions
%%%%%%%%%%%%%%%%%%%%%%%%%%%%%%%%%%%%%%%%%%%%%%%%%%%%%%%%%%%%%%%%%%%%%%%%%%%%%%
\clearpage

\noindent {\Large\bf{ Figure Captions }}\\ \\

\noindent {\bf Figure~\ref{fig.1}:} Relationship between stock price and trading activity.
Representative example of time series derived from the Trades and Quotes (TAQ) database for transactions of stock in Compaq Computer Corp. (CPQ) registered on the NYSE.
(a) Price of CPQ stock over a three week period from 20 Feb.- 8 Mar. 1996 (42606 trades). On 1 Mar. 1996 Compaq reported that it would cut product prices in order to meet sales targets, leading to a drop in the stock price. (b) Intertrade times (ITT) of CPQ stock over the same period. 
Data exhibit complex fluctuations, a daily pattern of trading activity (with short ITT at the open and close of a trading day and longer ITT in between),
and highly heterogeneous structure, as seen in the flurry of trades following the price drop. The relaxation time of the ITT response following the price drop extends over several days, suggesting that information may be contained in the temporal structure of trading activity. Data include transactions occurring between 9.30am and 4pm EST, excluding weekends and holidays.\\ \\

\noindent {\bf Figure~\ref{fig.2}:} Different correlation properties in intertrade times for stocks registered on the NYSE and NASDAQ markets.
(a) Root-mean-square fluctuation function $F(n)$ obtained using DFA-2 analysis, for the intertrade times (ITT) of stock in NASDAQ company Sun Microsystems (SUNW) and NYSE company Compaq Computer Corp. (CPQ). Here $n$ indicates the time scale in number of trades. We normalize the time scale $n$ by the daily average number of trades for each stock, so that a unit normalized scale indicates one trading day (marked by a dashed line). The scaling curves are vertically offset for clarity.
While both companies have similar market capitalisations, industry sectors and average levels of trading activity (average ITT) and exhibit long-range power-law correlations over a broad range of scales, the scaling behaviour of the intertrade times for the two stocks is quite different.
For CPQ we find a pronounced crossover from weaker correlations over time scales smaller than a day, to stronger correlations over time scales larger than a trading day ($\alpha_{2}^{ITT_{CPQ}} > \alpha_{1}^{ITT_{CPQ}}$).
In contrast, the scaling function $F(n)$ for SUNW does not exhibit such a crossover, and we find much stronger correlations over time scales smaller than a trading day compared with CPQ ($\alpha_{1}^{ITT_{SUNW}}>\alpha_{1}^{ITT_{CPQ}}$).
Correlation exponents $\alpha_{1}$ and $\alpha_{2}$ characterising the temporal structure in ITT for (b) one hundred NYSE stocks and (c) one hundred NASDAQ stocks, of companies with a broad range of market capitalisations and industry sectors.
Stocks are ranked in order of decreasing average value of ITT ($\overline{ITT}$) (as in Tables~\ref{Table 1} \& \ref{Table 2}), and are split into subsets (marked by vertical dashed lines) with matching $\overline{ITT}$ and approximately equal numbers of stocks.
We estimate $\alpha_{1}^{ITT}$ over scales from 8 trades to half of the daily average number of trades (for stocks with fewer than $1.5 \times 10^{5}$ trades/year), and to a third of the daily average number of trades (for stocks with more than $1.5 \times 10^{5}$ trades/year).
We estimate $\alpha_{2}^{ITT}$ over scales from 3 to 100 times the daily average number of trades.
Group averages and standard deviations of $\alpha_{1}^{ITT}$ and $\alpha_{2}^{ITT}$ are shown to the right of the panel for each market.
Systematically higher values of $\alpha_{1}^{ITT}$ for the NASDAQ stocks 
(statistically significant difference with p-value $p < 10^{-7}$ by the Student's t-test)
suggest
an underlying influence of market structure on the temporal organisation of intertrade times over scales within a trading day. In contrast, no systematic differences between the two markets are observed in the values of $\alpha_{2}^{ITT}$, characterising correlation properties of intertrade times over scales above a trading day ($p=0.03$ by the Student's t-test). We find similar results when we analyse trading activity at high resolution in terms of the number of trades per minute: a crossover at one trading day and stronger correlations for NASDAQ stocks compared with NYSE stocks over time scales less than a day (features which were not observed in previous studies \cite{Plerou00,Bonanno00}). We further observe an increasing trend in the values of $\alpha_{1}^{ITT}$ and $\alpha_{2}^{ITT}$ with decreasing $\overline{ITT}$ for both markets. \\ \\

\noindent {\bf Figure~\ref{fig.3}:} Comparing long-range correlations in ITT for groups of stocks with varying average levels of trading activity on the NYSE and the NASDAQ.
(a) Dependence of exponent $\alpha_{1}^{ITT}$, characterising the strength of correlations in ITT over scales from seconds up to a trading day, on the average level of trading activity. Each datapoint represents the group average over a subset of stocks, with a matching range of average intertrade times $\overline{ITT}$ for the two markets.
Stocks are grouped into subsets as indicated by vertical dashed lines in Fig.~\ref{fig.2}b,c. The consistent difference in the scaling exponent $\alpha_{1}^{ITT}$ between NYSE and NASDAQ stocks suggests that independent of company characteristics such as market capitalisation and industry sector, the temporal organisation of ITT within a trading day carries an imprint of market structure. 
(b) Dependence of exponent $\alpha_{2}^{ITT}$ characterising correlations in ITT over time scales from a trading day to several months, on the average level of trading activity. On both markets we observe similar behaviour with no systematic difference in the values of $\alpha_{2}^{ITT}$ between NYSE and NASDAQ subsets of stocks with matching ranges of $\overline{ITT}$. 
These results suggest that over time horizons longer than a trading day, the impact of market structure on trading dynamics is less pronounced as more information is available to investors over longer time scales, driving their trading activity. The resulting more coherent behaviour of investors is reflected in stronger correlations $(\alpha_{2}^{ITT} > \alpha_{1}^{ITT})$ over longer time scales. \\ \\

\noindent {\bf Figure~\ref{fig.4}:} Correlation properties of intertrade times of companies that moved from the NASDAQ to the NYSE during 1994-1995. (a) Fluctuation function $F(n)$, obtained using DFA-2 analysis on ITT of stock in the company Mid-Atlantic Medical Services Inc. while it was on the NASDAQ (3 January 1994 - 29 September 1994) and then after it moved to the NYSE (30 September 1994 - 30 November 1995). Here $n$ indicates the scale in number of trades and the vertical dashed lines indicate the average daily number of trades while on the NYSE or the NASDAQ. The two scaling curves are vertically offset for clarity. After the move to the NYSE there is a decrease in the correlation exponent at time scales within a trading day and a pronounced crossover to stronger correlations with a higher exponent at larger time scales. (b) Correlation exponent $\alpha_{1}^{ITT}$ characterising fluctuations over time scales less than a trading day in ITT of stock in ten companies that moved from the NASDAQ to the NYSE. Companies are ranked in order of decreasing $\overline{ITT}$ while on the NYSE (as in Table~\ref{Table 3}) and the scaling range for $\alpha_{1}^{ITT}$ is the same as for the hundred NYSE and NASDAQ stocks (Fig.~\ref{fig.2}b,c). For all companies there is a decrease in $\alpha_{1}^{ITT}$ after the move to the NYSE, indicating that the transition to weaker correlations in ITT over time scales less than a day is due to the NYSE market structure and not to company-specific characteristics. \\ \\

\noindent {\bf Figure~\ref{fig.5}:} Relation between correlations in intertrade times and stock price dynamics (a) Dependence of exponent $\alpha_{1}^{|RET|}$ characterising power-law correlations in absolute logarithmic price return fluctuations, on correlation exponent $\alpha_{1}^{ITT}$ characterising intertrade times within a trading day. Data represent one hundred NYSE (Table~\ref{Table 1}) and one hundred NASDAQ (Table~\ref{Table 2}) stocks. We calculate price returns over 1-minute intervals and $\alpha_{1}^{|RET|}$ over time scales from 8 to 180 minutes ($\approx$ half a trading day, which is 390 minutes). The positive relationship between $\alpha_{1}^{ITT}$ and $\alpha_{1}^{|RET|}$ indicates that stronger correlations in ITT are coupled with stronger correlations in price fluctuations.
This finding suggests that price fluctuations are not merely a response to short-term bursts of trading activity \cite{Jones94,Engle98}: rather the fractal organisation of price fluctuations over a broad range of time scales is linked to the observed underlying scaling features in the series of intertrade times.
Dependence of stock price volatility $\sigma^{RET}$ on  (b) the correlation exponent $\alpha_{1}^{ITT}$ and (c) the average value of ITT for the same stocks as in (a).
We calculate $\sigma^{RET}$ as the standard deviation of daily logarithmic price returns over six-month periods, averaging over all six-month periods throughout the entire record of each stock. Our results show no strong dependence between stock price volatility $\sigma^{RET}$ and average level of trading activity, rather the volatility appears sensitive to the strength of the temporal correlations in ITT.
These findings suggest that scale-invariant features in transaction times may play an important role in price formation. Furthermore, both dynamic and static properties of stock prices appear to be influenced by market-specific features in transaction timing: stronger power-law correlations in ITT (higher values of $\alpha_{1}^{ITT}$) for NASDAQ stocks are matched by stronger power-law correlations in price fluctuations (higher values of $\alpha_{1}^{|RET|}$) and higher volatility ($\sigma^{RET}$), compared with NYSE stocks.

\clearpage
%%%%%%%%%%%%%%%%%%%%%%%%%%%%%%%%%%%%%%%%%%%%%%%%%%%%%%%%%%%%%%%%%%%%%%%%%%%%%%%%
%   Figures
%%%%%%%%%%%%%%%%%%%%%%%%%%%%%%%%%%%%%%%%%%%%%%%%%%%%%%%%%%%%%%%%%%%%%%%%%%%%%%

\begin{figure}
\centering
\includegraphics[width=14cm]{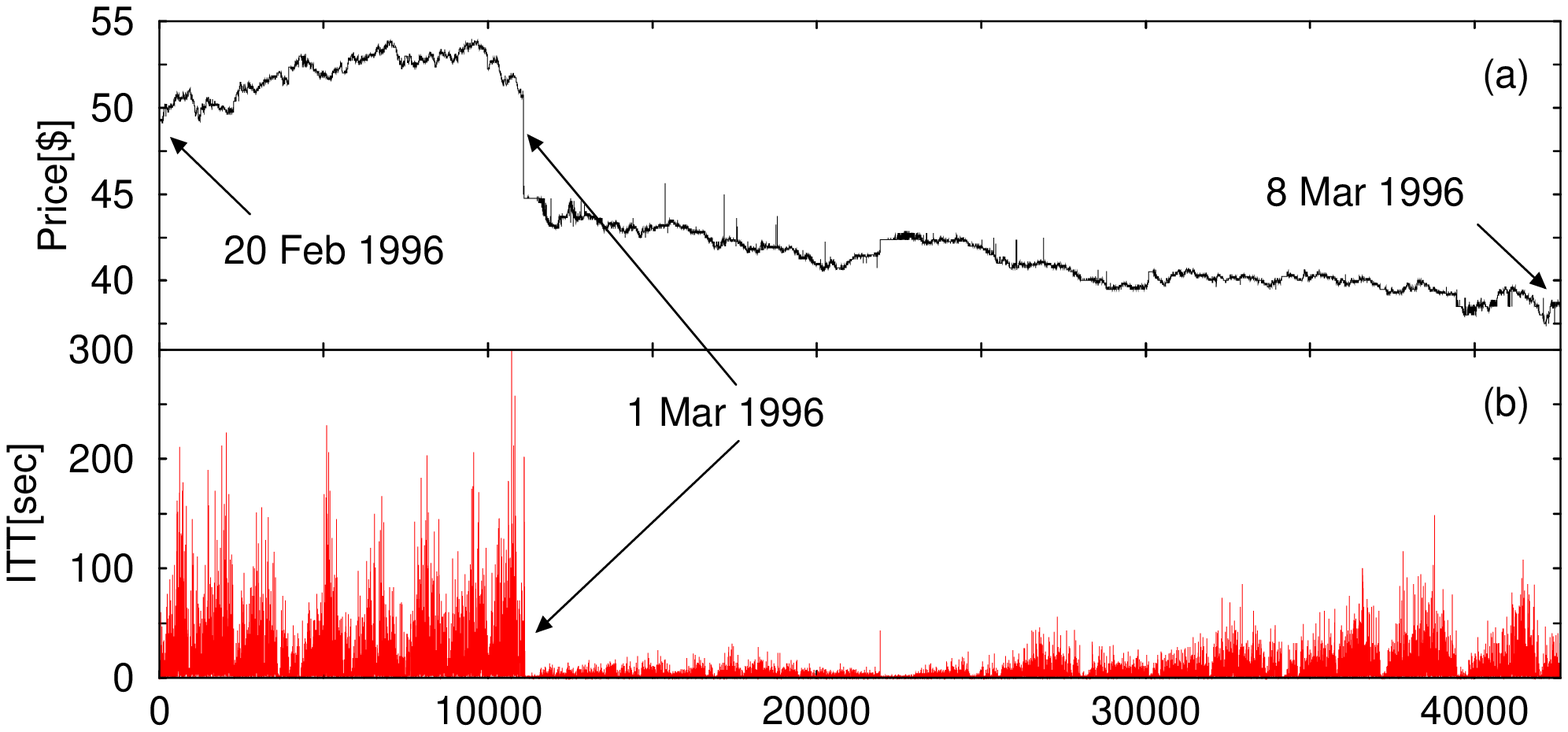}
\caption{ }
\label{fig.1}
\end{figure}

\begin{figure}
\centering
\includegraphics[width=8.6cm]{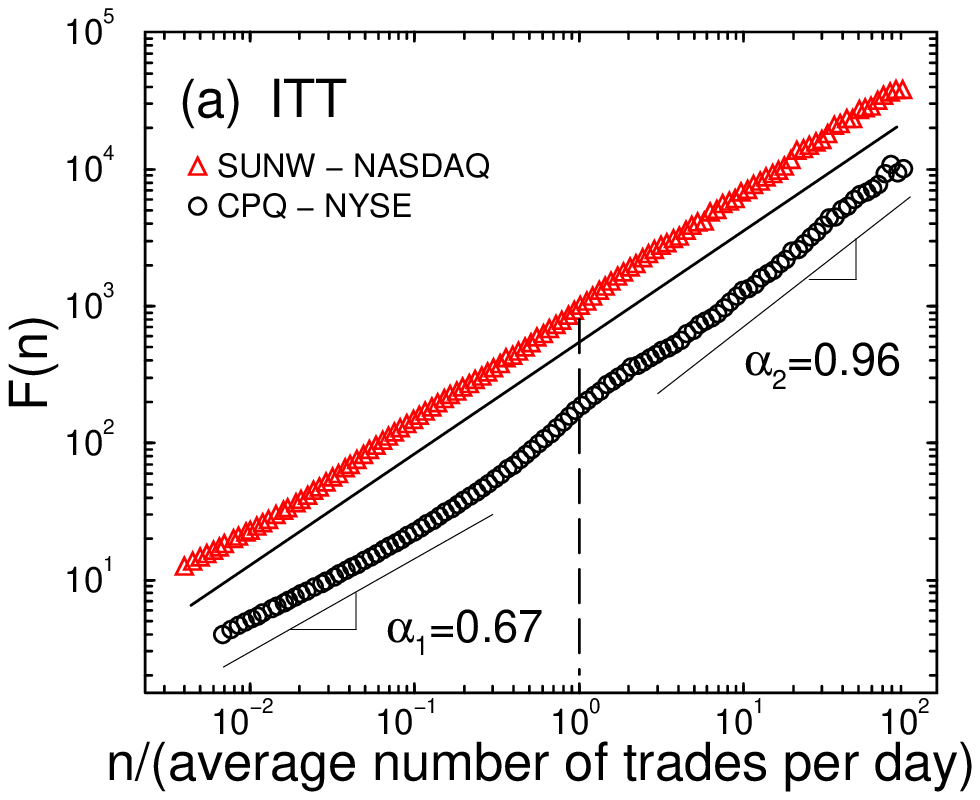}
\includegraphics[width=17.5cm]{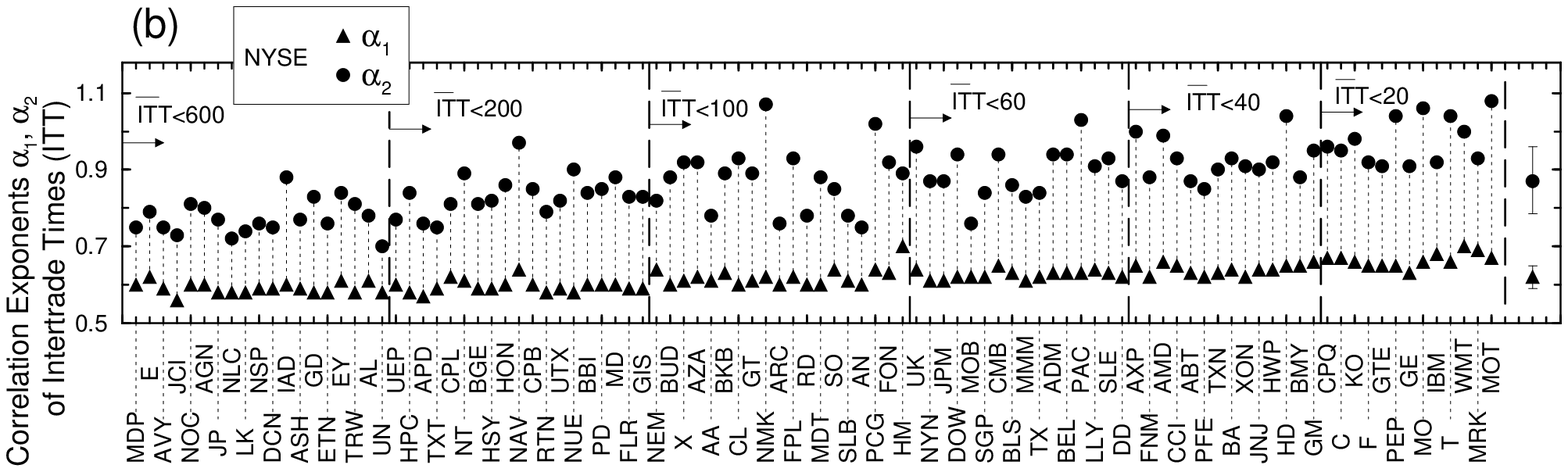}
\includegraphics[width=17.5cm]{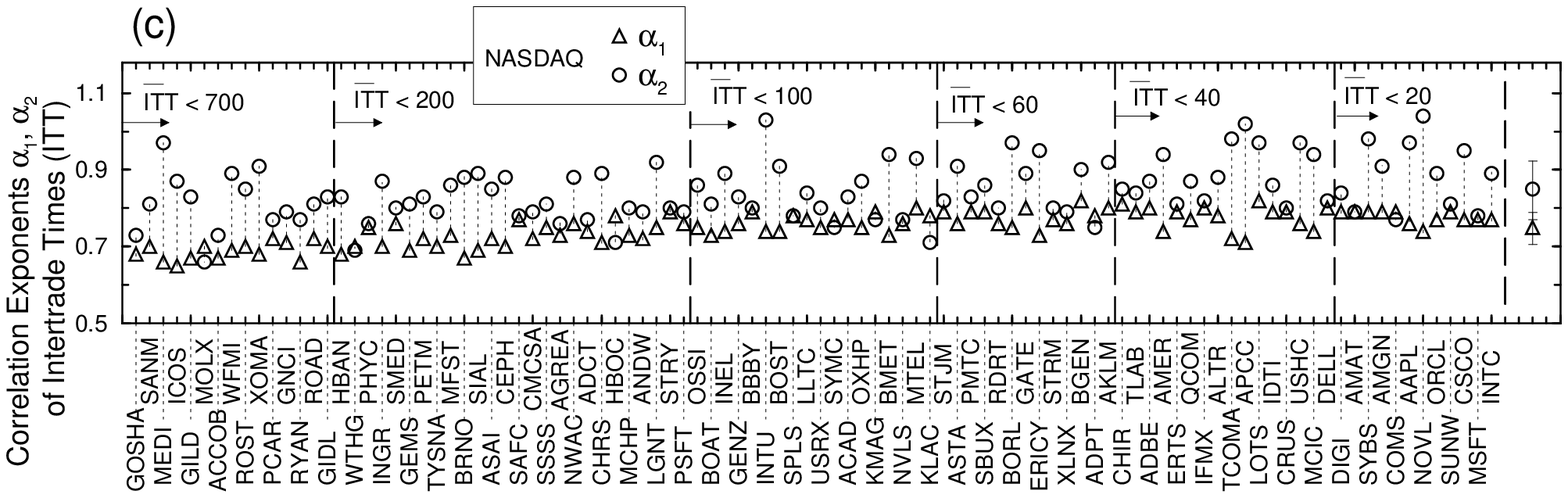}
\newpage
\caption{  }
\label{fig.2}
\end{figure}

\begin{figure}
\centering
\includegraphics[width=8.5cm]{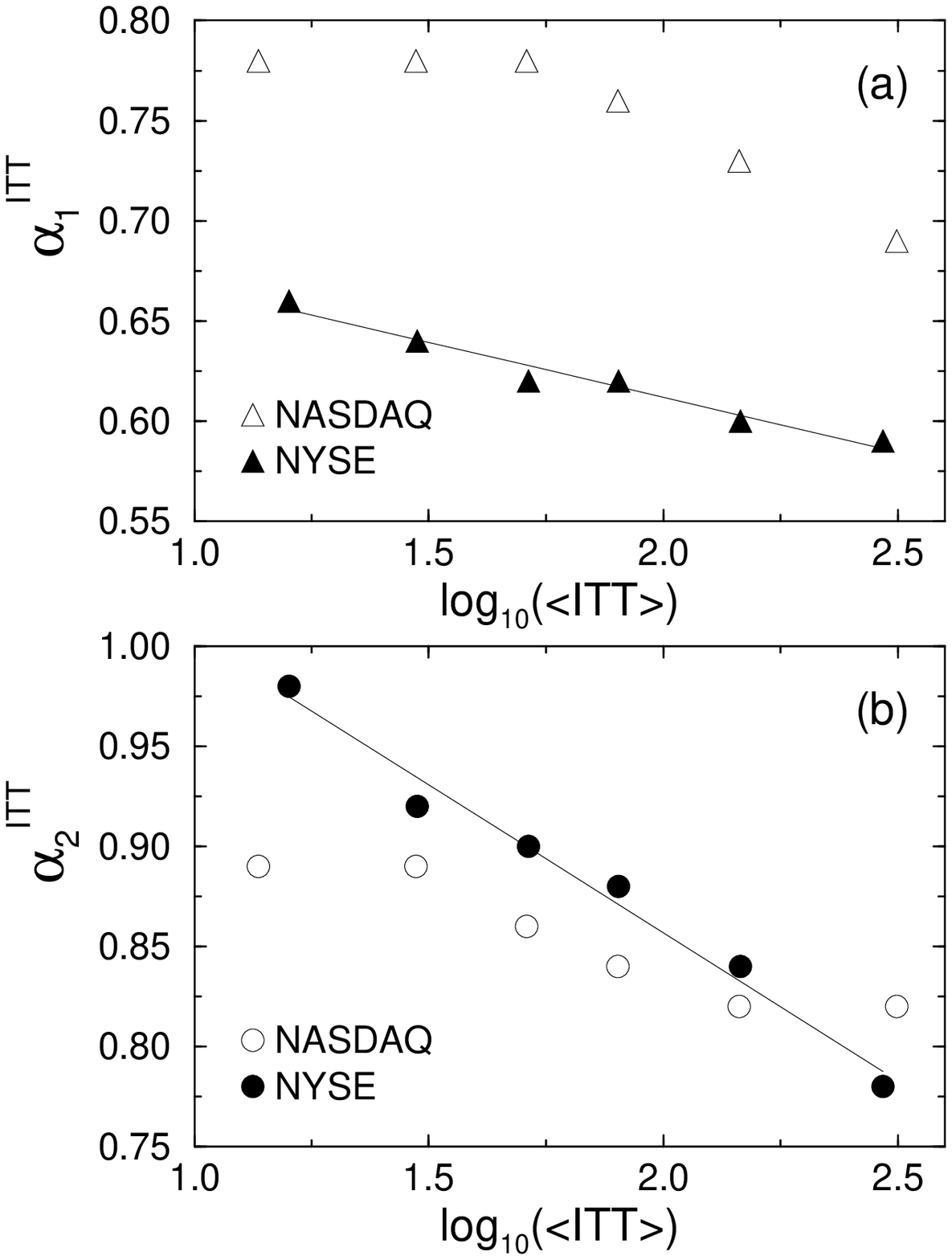}
\caption{ }
\label{fig.3}
\end{figure}

\begin{figure}
\centering
\includegraphics[width=8.6cm]{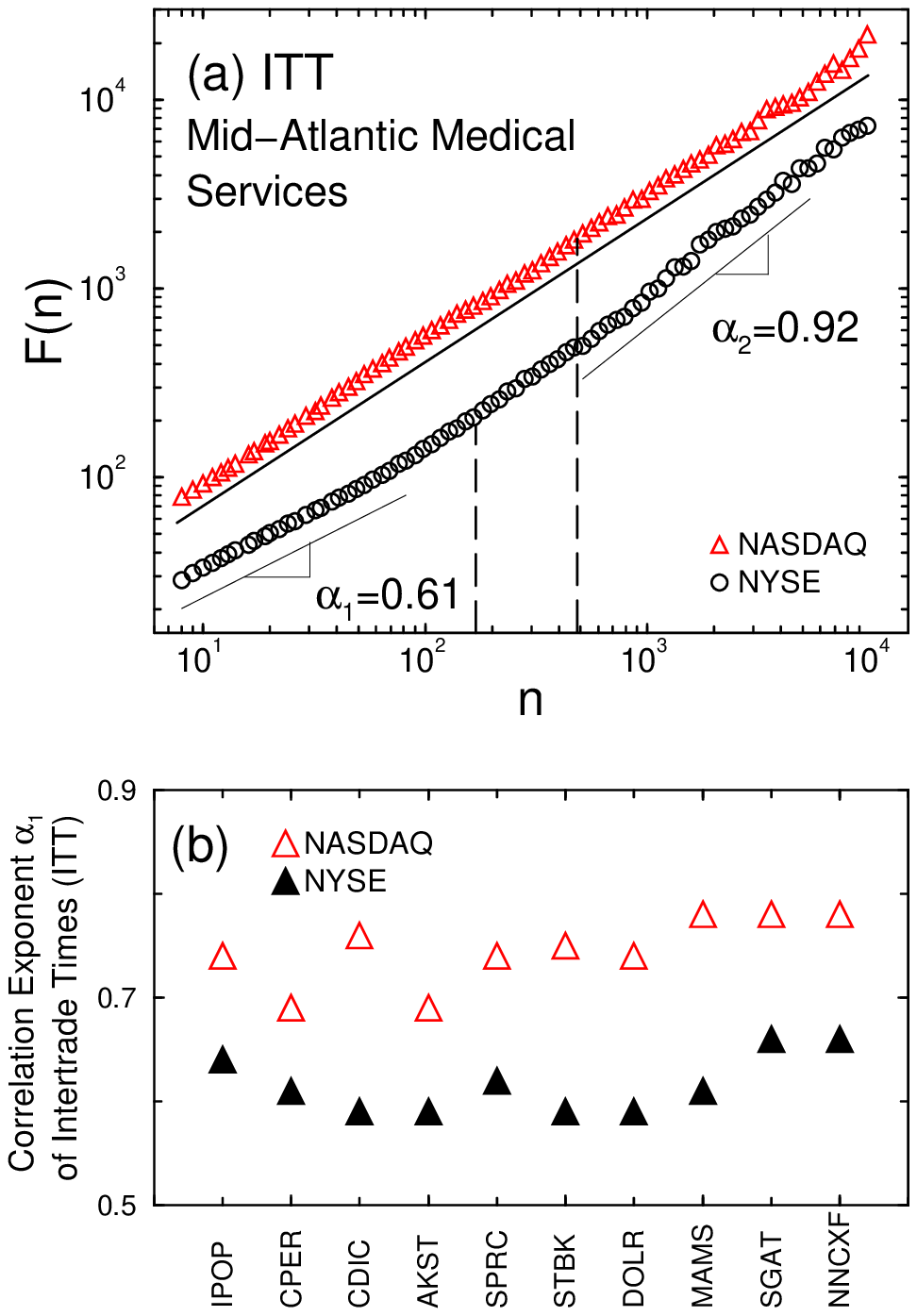}
\caption{ }
\label{fig.4}
\end{figure}

\begin{figure}
\centering
\includegraphics[width=8.5cm]{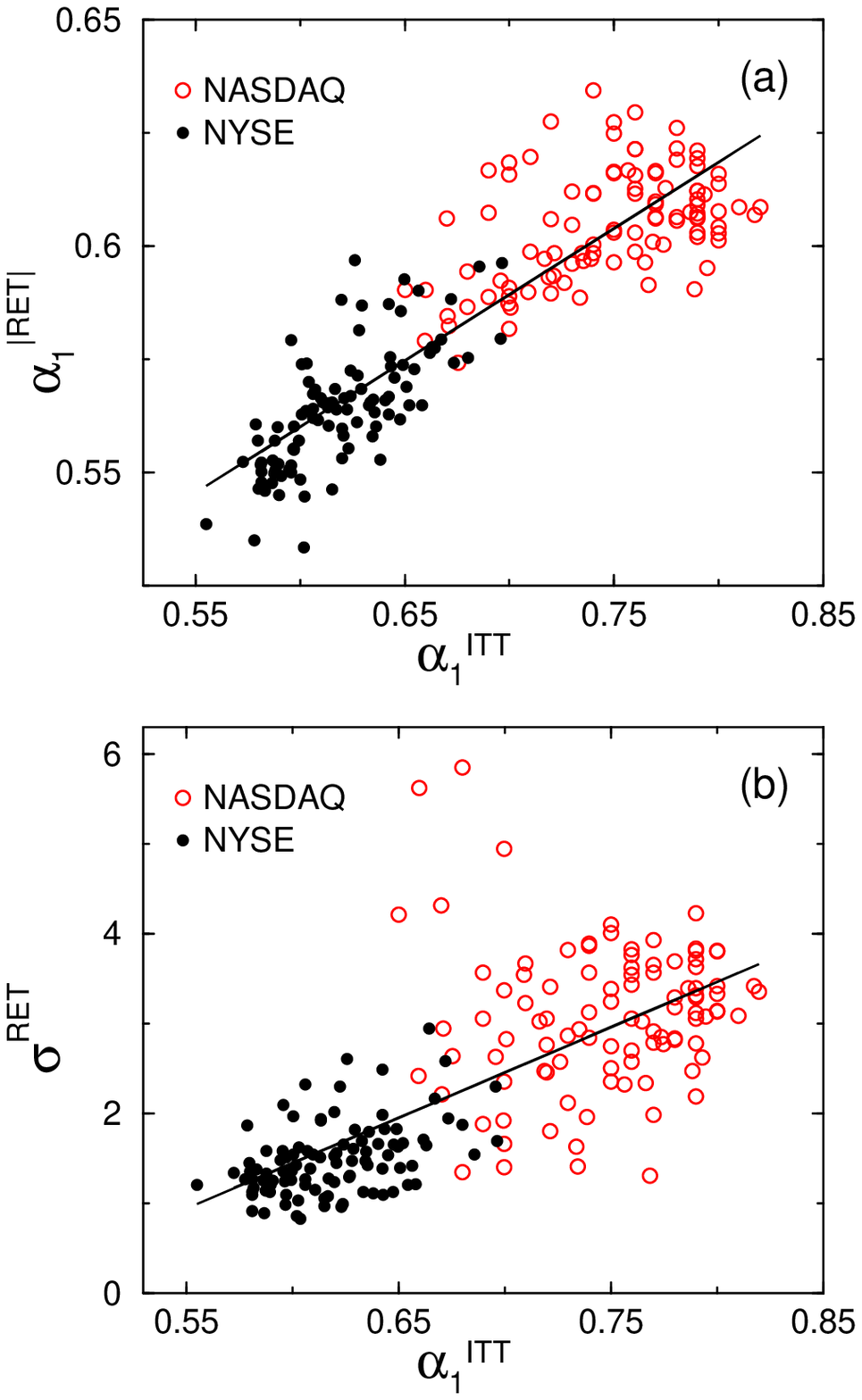}
\includegraphics[width=8.0cm]{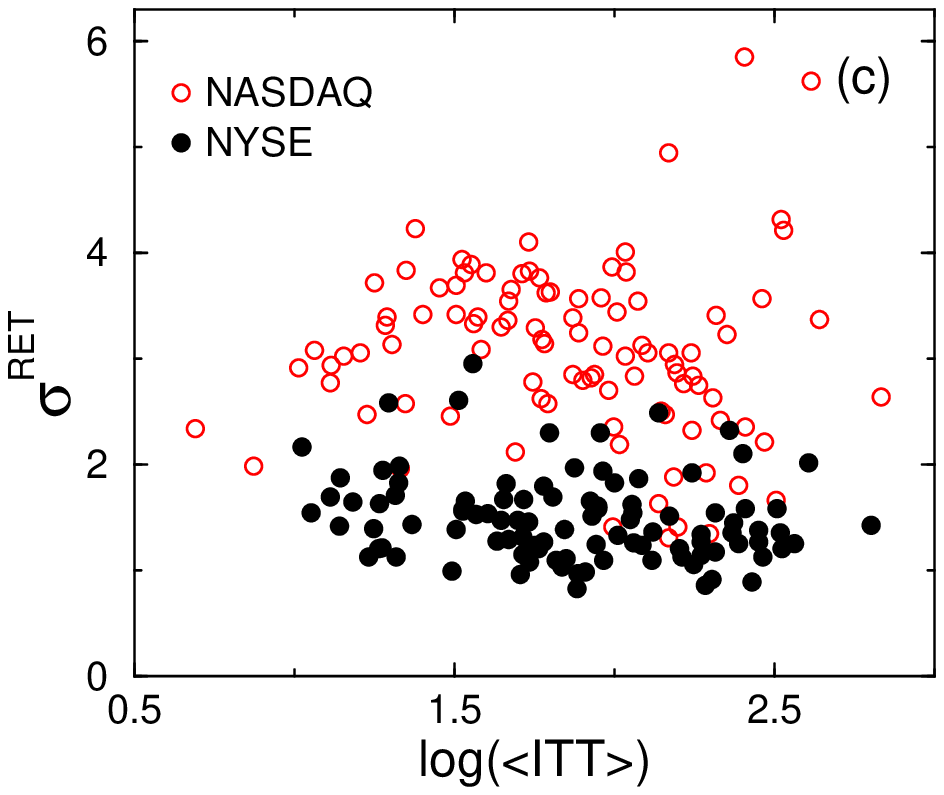}
\caption{ }
\label{fig.5}
\end{figure}

\end{document}